\documentclass[acmsmall]{acmart}
\usepackage{natbib}
\usepackage{amsmath,amsfonts}
\usepackage{algorithmic}
\usepackage{graphicx}
\usepackage{textcomp}
\usepackage{booktabs}
\usepackage{enumitem}
\usepackage{multirow}
\usepackage{xspace}
\usepackage{caption}
\usepackage{color}
\usepackage[table]{xcolor}
\usepackage{array}
\usepackage{subcaption}
\usepackage[most]{tcolorbox}
\usepackage[normalem]{ulem}

\usepackage{cleveref}

\setlist[itemize]{leftmargin=*}

\def\BibTeX{{\rm B\kern-.05em{\sc i\kern-.025em b}\kern-.08em
    T\kern-.1667em\lower.7ex\hbox{E}\kern-.125emX}}

\newcommand*{\mycode}{\ttfamily}

\newcommand*{\gpt}{GPT-4o-mini\xspace}
\newcommand*{\qwenlarge}{Qwen2.5-32B-Ins\xspace}

\newcommand*{\qwenlargeabbr}{Qwen-32B\xspace}
\newcommand*{\qwensmallabbr}{Qwen-7B\xspace}
\newcommand*{\dscoderabbr}{DS-Coder\xspace}

\tcbset{
    answerbox/.style={
        enhanced,
        colframe=black,
        drop shadow={opacity=0.5},
        boxrule=0.5pt,
        breakable,
        left=2pt,
        right=2pt,
        top=1pt,
        bottom=1pt,
    }
}

\setcopyright{cc}
\setcctype{by}
\acmDOI{10.1145/3832275}
\acmYear{2026}
\acmJournal{PACMSE}
\acmVolume{3}
\acmNumber{ISSTA}
\acmArticle{ISSTA184}
\acmMonth{10}
\acmSubmissionID{issta26main-p2106-p}
\received{2026-01-30}
\received[accepted]{2026-06-25}

\begin{document}

\title{What Makes In-Context Examples Effective for Code Generation?}

\author{Dongze Li}
\orcid{0009-0004-6482-8509}
\authornote{Co-first authors.}
\email{dongze.li@connect.ust.hk}

\author{Songqiang Chen}
\orcid{0000-0002-1220-8728}
\authornotemark[1]
\email{i9s.chen@connect.ust.hk}

\author{Jialun Cao}
\orcid{0000-0003-4892-6294}
\email{jcaoap@cse.ust.hk}
\authornote{Corresponding authors.}

\author{Shing-Chi Cheung}
\orcid{0000-0002-3508-7172}
\email{scc@cse.ust.hk}
\authornotemark[2]

\affiliation{%
    \institution{The Hong Kong University of Science and Technology}
    \city{Hong Kong}
    \country{China}
}
\affiliation{%
    \institution{Guangzhou HKUST Fok Ying Tung Research Institute}
    \city{Guangzhou}
    \country{China}
}

\begin{abstract}
In-Context Learning (ICL) has emerged as a promising solution to enhance the code generation capabilities of Large Language Models (LLMs) by incorporating code examples inside the prompt to let LLMs learn from demonstrations. 
However, despite their effectiveness gains, it remains unclear which specific properties of ICL-provided code examples (e.g., solution insight, essential contextual information, identifier naming styles, code formatting) drive these gains. 
This paper systematically investigates the impact of different sources and internal features of code examples on ICL for code generation through controlled experiments on contest-style programming questions and repository-level tasks.
Our results show that while LLMs struggle to extract generalizable problem-solving insights from provided solutions to similar questions or repository snippets, their retrieval-augmented ICL performance can significantly benefit from explicit contextual information, such as input/output demonstrations, required helper functions, and namespace information. 
Through targeted mutation operators, we further find that identifier naming is substantially more critical than code formatting or low-level implementation details, with the elimination of descriptive variable names causing performance drops of up to 30 percentage points.
Finally, we demonstrate that LLMs significantly prefer semantically meaningful identifier names and that adherence to surface-level naming conventions is far less important than semantic clarity.
These findings provide practical guidelines for constructing effective ICL code examples and highlight challenges in reflection-based learning for code generation.
\end{abstract}



\maketitle

\section{Introduction}

Code Generation is one of the most popular scenarios of AI-assisted software development.
Modern Large Language Models (LLMs) have demonstrated remarkable proficiency in code generation tasks, effectively producing functional code snippets that address natural language requirements~\cite{doi:10.1126/science.abq1158, qwen2025qwen25technicalreport, chen2021humaneval}.
Recent advancements leverage In-Context Learning (ICL)~\cite{llm-fewshotlearner}, where demonstration examples are embedded in the prompt alongside task descriptions, enabling LLMs to solve novel problems without parameter updates or fine-tuning.
This approach offers substantial advantages over traditional supervised fine-tuning methods~\cite{wang-etal-2021-codet5, feng-etal-2020-codebert}, which often require extensive computational resources and model parameter updating.
Instead, ICL elegantly combines selected demonstration examples with the original user requirements in the prompt, enabling the model to learn patterns and approaches from the incorporated human knowledge~\cite{DBLP:conf/acl/LuBM0S22, cot-prompting, wu-etal-2023-self}.
Various studies have validated the effectiveness of ICL with code examples for code intelligence tasks~\cite{acecoder, 10.1145/3715908LAIL, 10.1145/3551349.3559548, zhou2023docprompting}.

Substantial research focuses on identifying key factors that constitute effective ICL examples for code intelligence tasks.
Their findings include the dependency of LLM performance under ICL on the semantic similarity of code examples, as well as various factors affecting ICL performance, e.g., example selection~\cite{10.1109/ASE56229.2023.00109goodICLdemo}.
Although their insights help enhance code example-based ICL for code generation, an important question remains underexplored: {\bfseries \itshape Which specific properties (e.g., implicit solution logic, essential and required context, and descriptive identifier naming) in ICL code examples contribute most significantly to the observed performance improvements?}

Understanding the relative impact of individual code features may help inspire example selection and enhance LLM performance in code generation tasks, providing guidelines for ICL optimization and AI-assisted coding tasks.
To bridge this research gap, we conduct a fine-grained experiment that systematically manipulates both the source and internal structure of ICL code examples: 
(1) we vary how examples are retrieved and which kinds of contextual information they provide; 
(2) we selectively eliminate different code features via mutation operators; and 
(3) we synthesize diverse naming styles while preserving program behavior to probe LLMs' preferences on identifier naming.
Specifically, we answer the following three research questions (RQs) to achieve the goal:

\textbf{RQ1: What kind of information contributes most to ICL performance improvement?}
Since LLMs can, in principle, generalize their knowledge and capabilities to solve unseen tasks by reflection and inference~\cite{10.5555/3666122.3666499}, this research question examines whether insights from similar programming problems or relevant code snippets can improve code generation via ICL.
We focus on two scenarios widely studied in code generation research~\cite{jiang2026surveyLLM}: contest-level questions where task descriptions are complete and self-sufficient, and repository-level tasks that require external functions and contextual dependencies.
We construct retrieval-augmented ICL (RAG-ICL) systems for both scenarios, retrieve documents via multiple similarity metrics (sparse, dense, and tag-based retrievers), and embed the retrieved documents into prompts as ICL content.
Different from prior studies~\cite{10.1145/3717061, coderagbench} on retrieval-augmented code generation, which mainly focus on the overall effectiveness of different knowledge resources, we further investigate various types of information they provide (e.g., input/output demonstrations, required function and variable information, repository structural details). 
This setup allows us to evaluate not only whether LLMs can benefit from indirect problem-solving logic in retrieved examples, but also the key contexts bringing more significant contributions to ICL.

\textbf{RQ2: What kinds of code features contribute most to ICL performance?}
We investigate the impact of different code features by systematically eliminating each feature category from otherwise helpful ICL code examples.
Building on prior work, we categorize code features into naming information (e.g., local variable names, function signatures), format information (e.g., indentation, line separation, token layout), and implementation details (e.g., programming constructs, language-specific keywords).
We design a set of mutation operators that perturb each category while preserving the original program behavior as much as possible, and then measure how these manipulations affect ICL-assisted code generation on both LeetCode and DevEval.
Through controlled ablation studies, we quantify the relative importance of each feature category for code generation accuracy, semantic correctness, and overall quality metrics.

\textbf{RQ3: What naming characteristics of code align with LLMs' preferences?}
Existing works have indicated that different identifier naming styles have a significant effect on LLMs' comprehension of code examples~\cite{7503707improving_textual, codepromptzip}.
We further examine how different naming styles, ranging from fully symbolic to concise, verbose, or noisy, affect ICL outcomes.
We synthesize multiple naming variants for the same program (e.g., ordered symbolic names, concise but semantically precise names, overly verbose names, and high-frequency but semantically misleading names) while preserving program behavior.
This enables us to disentangle the effects of semantic informativeness, length, and adherence to naming conventions, and to identify naming patterns that most effectively support LLM understanding and generalization.

These three research questions collectively investigate the effective components of ICL code examples through progressively deeper analysis—from high-level retrieval strategies to fine-grained code-level features. RQ1 explores different retrieval techniques and codebase sources, establishing the overall landscape of ICL effectiveness.
RQ2 zooms into the code examples themselves, investigating which internal components (e.g., contextual information vs. implementation logic) contribute most to performance gains. 
RQ3 further investigates the effect of different variable naming styles, which have been found to affect LLMs on code comprehension tasks~\cite{10.1109/ASE63991.2025.00250}.

Our study results yield several interesting findings, which reveal implications for the effective usage of ICL for code generation and highlight the limitations of current LLMs in leveraging implicit information in ICL code examples.
First, we find that ICL contents explicitly supplying essential contextual information, such as input/output demonstrations, required function and variable information, or repository structural details, bring measurable gains for LLM code generation.
Nevertheless, LLMs struggle to generalize problem-solving insights from solutions to similar questions, indicating limited capability in reflection-driven ICL.
In addition, we observe that identifier naming in code, particularly the names of local variables, has a much larger effect on ICL-assisted performance than code formatting or low-level implementation details: eliminating descriptive variable names can cause a performance decrease of up to 30 percentage points.
Moreover, our experiments reveal that LLMs prefer semantically meaningful identifier names, while noisy and misleading names may cause significant performance degradation, especially in Python.
This suggests that semantic clarity, rather than length or adherence to surface-level conventions, is what most effectively supports model understanding and generalization.
These findings inspire several actionable insights for improving ICL for code generation: practitioners should prioritize semantically informative naming and the inclusion of essential contextual elements when selecting or constructing ICL examples, and the observed limitations in reflection call for further research on enhancing model reasoning capabilities through more abstract or structural representations.

Overall, this paper makes the following contributions:
\begin{itemize}
    \item We conduct a systematic investigation of how different sources and structures of code examples affect the effectiveness of code demonstration-based ICL, including retrieval-augmented ICL and controlled code example modifications, yielding generalized observations and actionable insights for ICL code example construction.
    
    \item We design and implement an example retrieval system equipped with diverse retrieval configurations, as well as a controlled code feature elimination framework, using carefully designed mutation operators to selectively remove key features regarding code naming, format, and implementation details from ICL code examples and modify naming styles, enabling fine-grained performance attribution of effective ICL-provided code examples.

    \item We identify explicit contextual information (e.g., input/output demonstrations, required function/variable information, repository structure) and meaningful identifier naming as the most beneficial properties among the studied ICL components, and recommend incorporating such essential contextual elements and emphasizing semantic clarity in naming when selecting or constructing ICL code examples to maximize LLM performance in code generation.
\end{itemize}

\section{Background \& Motivation}

\subsection{In-Context Learning}
ICL refers to the ability of LLMs to learn from a few examples in the form of demonstrations embedded as prompt context~\cite{llm-fewshotlearner, dong-etal-2024-survey}.
As illustrated in \Cref{fig:prompt-template}, a demonstration section is incorporated in addition to the original requirement.
By embedding task-relevant examples in the demonstration section, ICL enables LLMs to generalize and perform downstream tasks more effectively without updating model parameters.
Specifically, an ICL system maintains a space of demonstration examples \(\mathbb{D} = \{(d_i)\}\), where \(d_i\) represents a document containing the knowledge content~\cite{coderagbench, asai2024selfrag}.
When receiving a query, it first selects \(N\) relevant and potentially helpful documents for the user query from the demonstration example space,  then concatenates the examples with the query into a prompt for the LLM and finally instructs the LLM to further extract the required information and generate the answer.

\subsection{Motivation} \label{subsec:motivation}
ICL has demonstrated strong performance in both natural language processing and code intelligence tasks~\cite{acecoder, 10.1145/3715908LAIL, 10.1145/3551349.3559548, zhou2023docprompting}, prompting increasing interest in understanding the underlying factors that drive its effectiveness.
Recent studies have examined several factors contributing to ICL performance in code generation, e.g., example selection strategies, semantic similarity, and supervision methods~\cite{liu-etal-2022-makes, 10.1109/ASE56229.2023.00109goodICLdemo, patel-etal-2024-evaluating}.
More recently, \citet{coderagbench} demonstrate that directly incorporating the correct solution code as ICL content substantially enhances LLM performance, while retrieved examples from different knowledge sources (e.g., programming solutions, GitHub repositories) yield limited degrees of benefit.

However, two critical questions remain underexplored.
First, existing studies~\cite{10.1145/3717061, coderagbench} primarily evaluate different knowledge \textit{sources} for retrieval-augmented ICL, but do not systematically analyze which \textit{types of properties and information} within retrieved content (e.g., input/output demonstrations, required function/variable information, repository structure) are most beneficial.
Second, most prior work treats code examples as monolithic units, without disentangling the contributions of individual code \textit{features} such as identifier naming, formatting, and implementation details.

Understanding both dimensions, i.e., the essential information and the key code features, may enable practitioners to craft more targeted and informative demonstrations for code intelligence tasks.
Motivated by this, we conduct a systematic empirical study addressing both aspects: (1) we investigate which types of contextual information in retrieved examples contribute to performance gains, and (2) we perform a controlled ablation of distinct code features within ground truth examples to evaluate their individual impact on ICL performance.

\section{Study Design}

\subsection{Overall Pipelines}
\label{sec:overview}

We conduct an extensive empirical study to fill the gap in understanding useful properties in ICL code examples, as mentioned in \Cref{subsec:motivation}.
To cover different scenarios in code generation, we select two widely-used benchmarks as subjects:

\begin{itemize}
    \item LeetCode Questions: We adopt LeetCode questions from the LiveCodeBench benchmark~\cite{livecodebench} as our target task.
    These questions exhibit self-contained, algorithmic characteristics and are typically addressable through concise, standalone code implementations.
    \item DevEval Tasks: A real-world repository-level code generation benchmark containing 115 repositories covering 10 popular domains~\cite{deveval}.
\end{itemize}

The selected benchmarks represent two programming scenarios.
LeetCode questions are \emph{self-contained}: the prompt fully specifies the task, and solutions generally rely only on the standard library.
DevEval tasks, by contrast, are grounded in real repositories and may require \emph{external dependencies}, such as calling existing project modules/APIs not included in the task description.
These two scenarios (competition and repository-level) are summarized as representative dimensions of code generation benchmarks by~\citet{jiang2026surveyLLM}.

For a clean evaluation, we exclude tasks whose ground truth answers failed on the provided test cases from both benchmarks.
We adopt 362 out of 381 LeetCode questions in LiveCodeBench-v5 (from May 2023 to January 2025) and 1427 out of 1825 DevEval tasks.

We choose Python and Java as the studied programming languages.
Python is the officially supported programming language for code generation tasks in LiveCodeBench and DevEval. 
To investigate the effectiveness of ICL code examples across programming languages, we included Java for LeetCode questions due to its prevalence in real-world development and academic instruction, as supported by the statistics from IEEE Spectrum~\cite{ieee-spectrum} and the developer survey from Stack Overflow~\cite{stackoverflowsurvey}.
We carefully design the prompt template and build the execution environment for Java, following the implementation for Python in LiveCodeBench.

In RQ1, we construct RAG systems on two evaluated code generation benchmarks to explore the effective information automatically retrieved by existing widely-used RAG systems.
This helps to understand the impact of ICL code examples in code generation tasks under various settings, e.g., target tasks, knowledge databases, and retrieval methods~\cite{10.1145/3637528.3671470}.
For LeetCode questions, we collect all public LeetCode questions, as well as their correct solutions available in open-source datasets\footnote{Python Solution: \url{https://github.com/kamyu104/LeetCode-Solutions}; Java Solution: \url{https://huggingface.co/datasets/allenhung1025/leetcode}.}.
For DevEval tasks, we extract all functions, methods, and classes in the programming repositories as the retrieval database, following the practice of \citet{coderag-jiali}.
We then retrieve the most relevant documents based on representative similarity metrics~\cite{coderagbench} between the target task description and the documents in the database.
Subsequently, we embed both the original query and the corresponding documents into the ICL content.

In RQ2 and RQ3, we explore the key features in ICL code examples that enable LLMs to generate correct code.
We select \textit{Ground Truth} configuration for LeetCode questions and \textit{Embedding Models} configuration for DevEval, where LLMs show remarkable performance improvement via ICL, then eliminate certain features from the ICL-provided code examples using pre-designed mutation operators to formulate various ICL examples.
Through performance comparison on examples with specific features eliminated, we investigate the impact of each feature in ICL code examples.

\begin{figure}[tbp]
    \centering
    \captionsetup[sub]{skip=0.5pt}
    \setlength{\abovecaptionskip}{0pt}
    \setlength{\belowcaptionskip}{0pt}
    \begin{subfigure}[b]{0.49\textwidth}
        \centering
        \includegraphics[width=.88\textwidth]{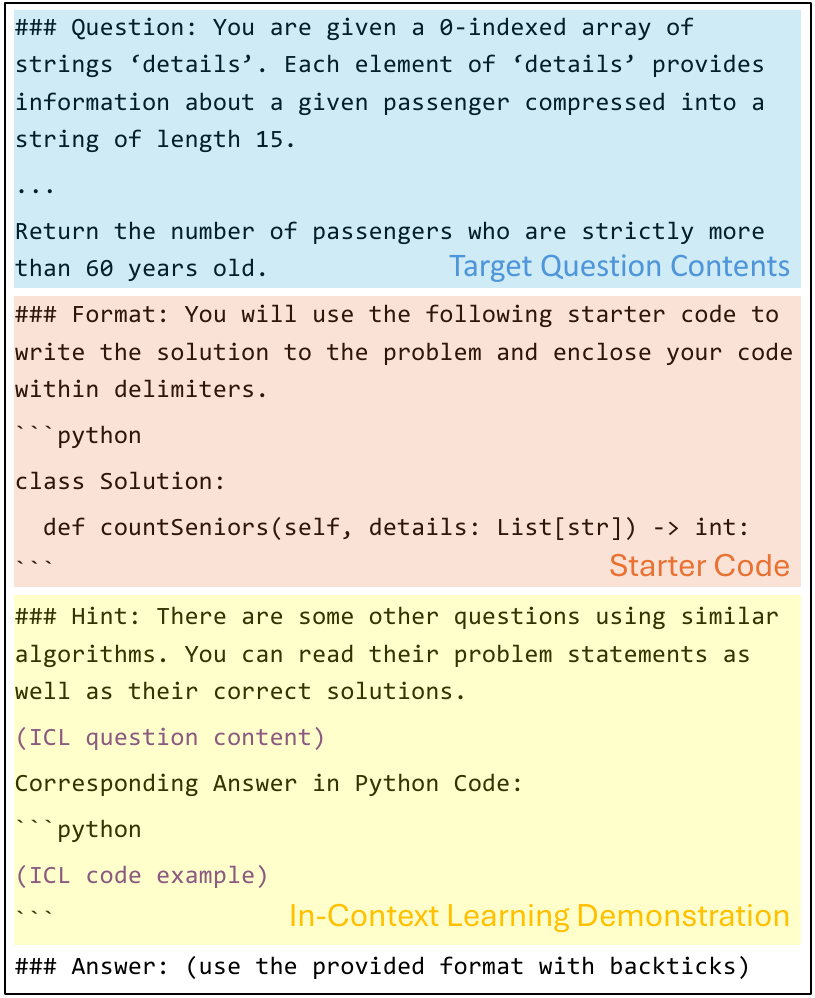}
        \label{fig:leetcode-prompt-template}
        \caption{LeetCode Prompt Template}
    \end{subfigure}
    \begin{subfigure}[b]{0.49\textwidth}
        \centering
        \includegraphics[width=.88\textwidth]{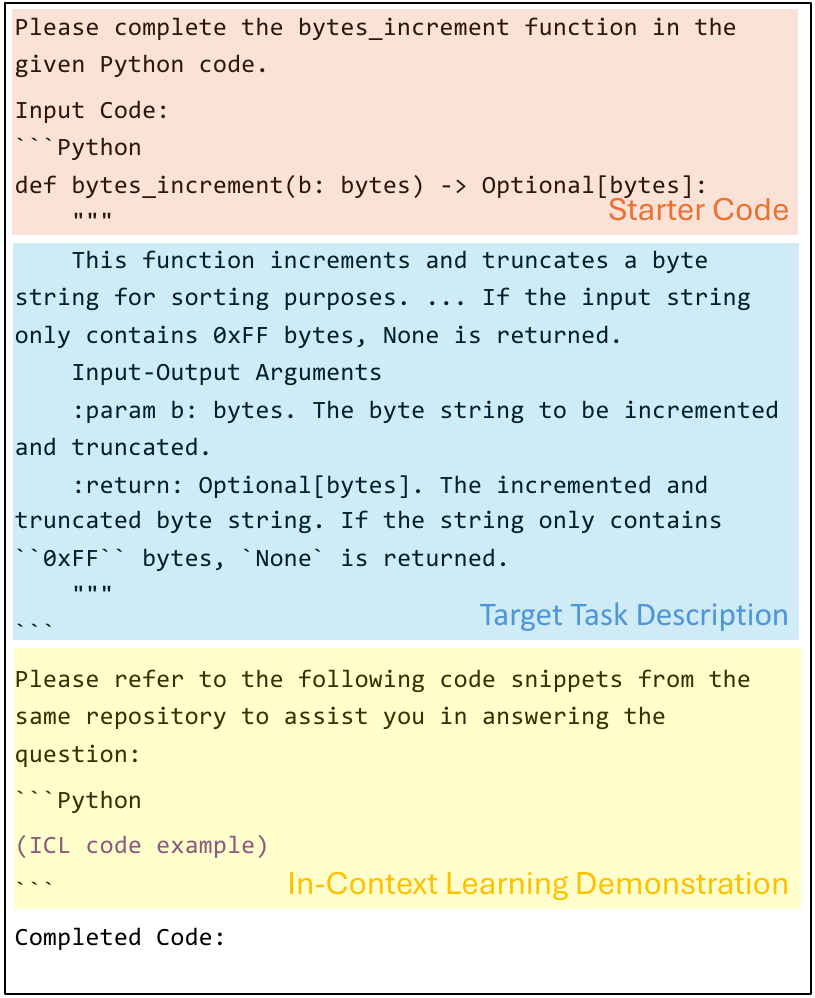}
        \label{fig:DevEval-prompt-template}
        \caption{DevEval Prompt Template}
    \end{subfigure}
    \caption{Examples of ICL-Enhanced Prompt for LeetCode and DevEval. \textit{Target Question Contents/Task Description} and \textit{Starter Code} parts follow the official prompts of LiveCodeBench and DevEval. \textit{ICL Demonstration} refers to \textit{(Question Content, Corresponding Answer)} for LeetCode and extracted code snippets for DevEval.}
    \label{fig:prompt-template}
\end{figure}

\subsection{Experimental Design}

In this section, we discuss the experimental design for each research question, including the RAG system setup for RQ1 and the mutation operators in RQ2 and RQ3.

\subsubsection{Experimental Design for RQ1 on Effective ICL-Incorporated Information}

\begin{figure}
    \centering
    \setlength{\abovecaptionskip}{1pt}
    \setlength{\belowcaptionskip}{0pt}
    \includegraphics[width=0.90\linewidth]{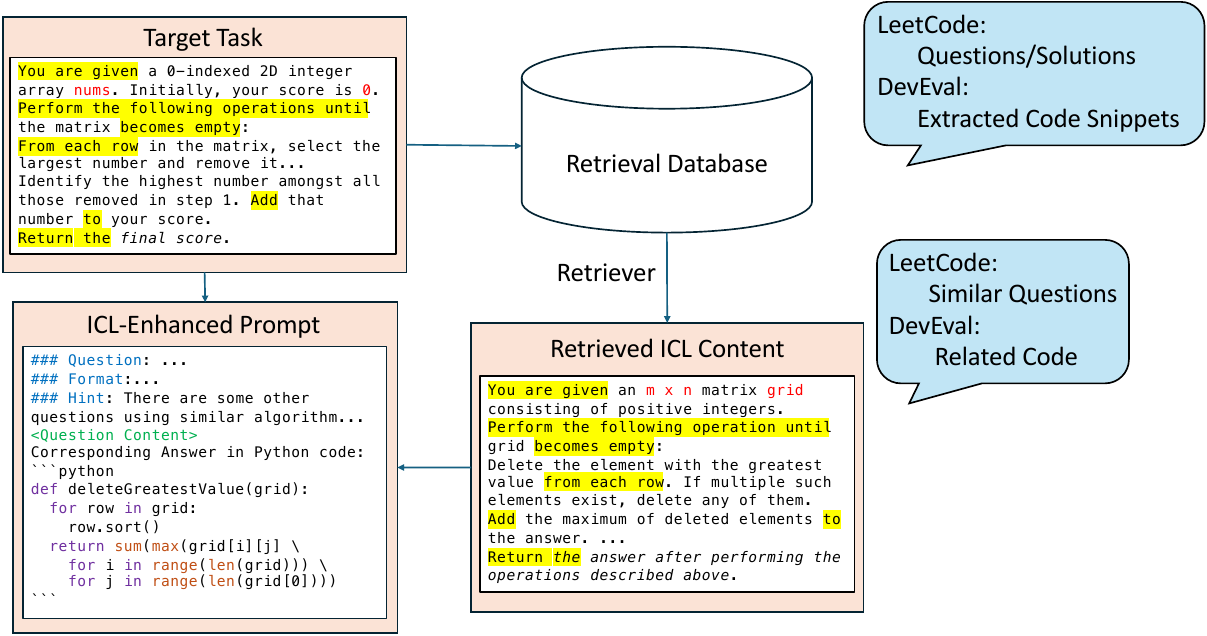}
    \caption{RAG System for RQ1. Documents from the retrieval database (Question/Solution for LeetCode and Related Code Snippets for DevEval) are retrieved and included in the prompt as the ICL example.}
    \label{fig:rq1-pipeline-leetcode}
\end{figure}

Recent studies have demonstrated remarkable successes of ICL in code generation tasks~\cite{10.1145/3715908LAIL, 10.1109/ASE56229.2023.00109goodICLdemo, su-etal-2024-evor, zhou2023docprompting}.
However, ICL is not a universally beneficial solution, as reported by~\cite{coderagbench, maekawa-etal-2024-retrieval, 10.5555/3692070.3694253selective}, indicating that ICL's effectiveness remains context-dependent.
Thus, in this section, we investigate how incorporating retrieved code examples as ICL demonstrations affects LLM performance under different retrieval and task configurations.

Specifically, we construct a RAG system~\cite{10.5555/3495724.3496517} that retrieves relevant documents from a database and embeds them into the prompt as ICL examples to assist the LLM in solving the target task.
For LeetCode questions, the RAG system retrieves documents defined as (\textit{Similar Question}, \textit{Correct Answer}) pairs, as illustrated in \Cref{fig:rq1-pipeline-leetcode}.
This setup simulates a typical programming problem demonstration, where correct solutions to related questions are expected to offer algorithmic and procedural insights~\cite{zhou2023leasttomost}.
Such implicit information may enable LLMs to reflect on problem-solving patterns and infer the correct approach for the target question.
For DevEval tasks, the RAG system follows the procedure of \citet{coderag-jiali}, retrieving individual code snippets (functions, methods, and classes) from programming repositories as candidate documents.
Specifically, we implemented a static code snippet extraction tool using the open-source library tree-sitter~\cite{brunsfeld2018treesitter}.
This tool parses the source code files into Abstract Syntax Trees (ASTs) and identifies class definitions and function definitions based on predefined node types corresponding to functions, methods, and classes.
It then extracts the code snippets and stores them into a RAG database with metadata (e.g., file path and line numbers) for efficient retrieval.
This tool is included in our artifact for reproducibility and further research use.
These retrieved snippets are expected to provide contextually relevant information, e.g., predefined helper functions, class attributes, and method implementations.

Following the established practices~\cite{asai2024selfrag, 10.1109/ICSE48619.2023.00179skcoder, ragstudyjy}, we retrieve the top five documents from the database (annotated as \(k{=}5\)) for each query using the following retrievers:
\begin{itemize}
    \item \textit{BM25}~\cite{10.1561/1500000019}: A popular sparse retriever that ranks documents based on term frequency and inverse document frequency (TF-IDF), known for its robustness in domain adaptation \cite{coderagbench}.
    \item \textit{Embedding Models}: We build dense retrievers based on the following open-source embedding models: gist-large~\cite{gistembed}; jina-v2-code~\cite{jinaembeddings}; codesage-small~\cite{DBLP:conf/iclr/ZhangATDNR0X24}, aligned with the retriever selection of CodeRAG-Bench \cite{coderagbench}. 
    These models encode text into dense vector representations, enabling semantic similarity-based retrieval.
    \item \textit{Algorithm Tag}: We collect the algorithm tags annotated by LeetCode for each question and retrieve other questions with the same algorithm tags. 
    This retriever helps reveal whether LLM can learn the implementation of the algorithm required by the target question from the solution code of other questions.
    Note that this retrieval method is not available for DevEval tasks.
    \item \textit{Random}: Following existing work~\cite{3694997instructive-code-retriever}, we include randomly retrieved documents as a baseline to isolate the effect of ICL content quality.
\end{itemize}

To simulate a common ICL scenario, we exclude ground-truth documents from the retrieval corpus. 
For LeetCode, we recorded {\ttfamily question\_id} for each solution for identification and removed the solution with the same ID as the target question from the retrieval candidates, preventing any presence of direct answers.
For DevEval, we recorded the file paths and line numbers of code snippets in the RAG codebase and the target function body (i.e., the ground truth) for each task.
During the retrieval process, we excluded all code snippets that overlapped with the position of the target function, i.e., in the same file and sharing the same line numbers.
This prevents inadvertent data leakage and ensures retrieved examples represent only external knowledge available to the model.
The retrieved ICL examples are then included in the prompt in the order of their similarity scores calculated by the retriever with the most similar example placed closest to the target question, as this ranking strategy is reported to be generally more effective compared to random or reversed ordering~\cite{10.1109/ASE56229.2023.00109goodICLdemo}.

We include two baseline setups to reflect the effect of ICL with different retrievers:
\begin{itemize}
    \item \textit{Ground Truth (GT)}: Incorporates the ground truth for the target task directly, which is found meaningful to reveal the upper-bound performance of ICL when the database is ideally broad and adequate \cite{coderagbench}.
    \item \textit{Zero Shot (ZS)}: Provides the target task and starter code shown in \Cref{fig:prompt-template} only, without any ICL demonstration examples, indicating LLMs' native parametric capabilities.
\end{itemize}

\newcommand*{\foe}{Function Ordered Erosion\xspace}
\newcommand*{\voe}{Variable Ordered Erosion\xspace}
\newcommand*{\fve}{Function \& Variable Ordered Erosion\xspace}
\newcommand*{\inr}{Indentation Removal\xspace}
\newcommand*{\lsr}{Line Separation Removal\xspace}
\newcommand*{\tok}{Tokenization\xspace}
\newcommand*{\nl}{Natural Language Representation\xspace}
\newcommand*{\lt}{Loop Transformation\xspace}

\subsubsection{Experimental Design for RQ2 on Effect of Different Code Features}

We conduct a controlled ablation study to investigate the effect of each code feature on ICL separately.
Specifically, we design systematic mutation operators to eliminate code features in different aspects that potentially contribute to the ICL performance.
Prior works categorize code features into three dimensions: \textit{code naming}~\cite{7503707improving_textual}, \textit{code format}~\cite{10.1145/1985441.1985454}, and \textit{implementation details}~\cite{10.1145/3790101}.
Naming features pertain to developer-assigned names within the code that describe its functionality, such as variable names, function names, and class names.
Format features refer to code format characteristics that distinguish code from plain text (e.g.,~indentation information and line separation). 
Implementation details encompass language-specific constructs and logic patterns, such as reserved keywords, e.g., {\mycode for}, {\mycode while}, and control flow structures.

\begin{figure}
    \centering
    \includegraphics[width=.80\linewidth]{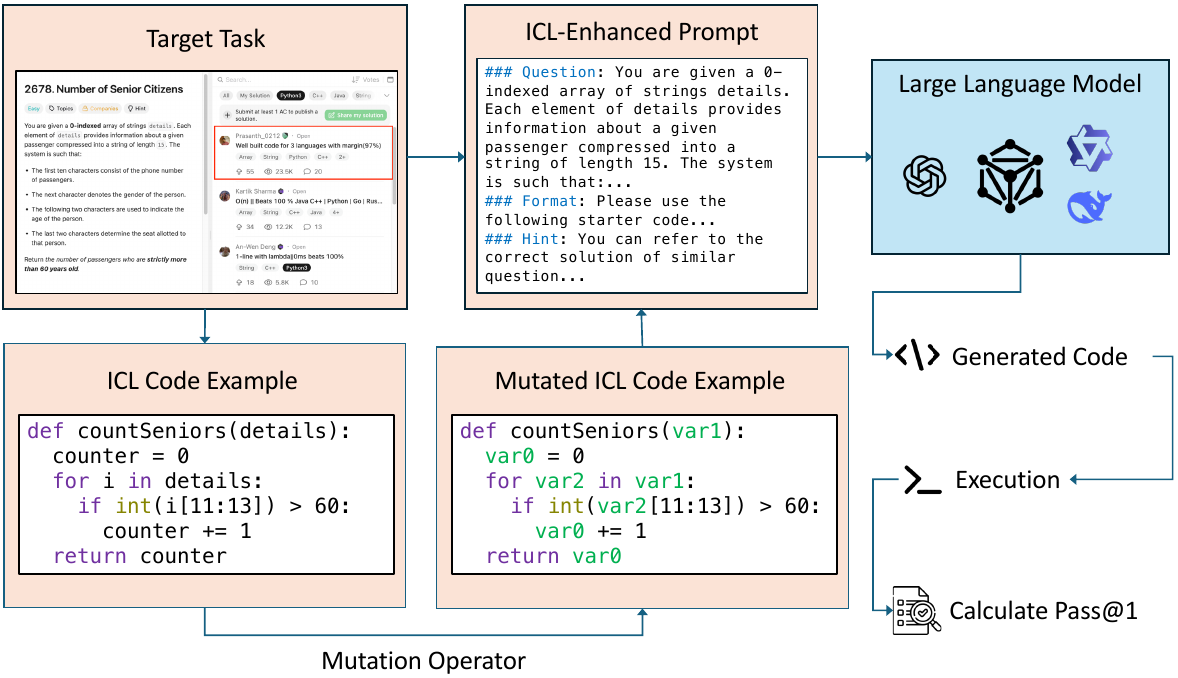}
    \setlength{\abovecaptionskip}{0pt}
    \setlength{\abovecaptionskip}{0pt}
    \caption{Pipeline to Apply Mutation Operators on ICL Example for RQ2 and RQ3.}
    \label{fig:pipeline-rq23}
\end{figure}

To isolate the impact of each feature category, we follow \citet{10.1145/3691620.3695072poorcodesumeval} to apply targeted mutation operators that alter specific aspects of the code. 
We design eight mutation operators.
The illustration of different mutation operators applied to the example in \Cref{fig:pipeline-rq23} is available in our artifact~\cite{artifact}.

\textbf{Naming Features}: These mutation operators preserve original code runtime behavior while renaming local variables, function parameters, and function names that describe their functionality.
We conducted a sanity check to ensure runtime behavior preservation by directly executing the mutated solution on the original tests.

\begin{itemize}
    \item \textit{\foe (FOE)}~\cite{10.1145/3428293foe}: Renames functions and class methods using ordered symbolic notation, e.g., {\mycode function0}, {\mycode function1} as well as corresponding function calls, while retaining built-in functions unchanged, e.g., {\mycode print}, {\mycode map}.
    For LeetCode, the entry function of the original code is renamed to {\mycode entry\_function} to indicate the entry point to LLMs.
    \item \textit{\voe (VOE)}: Renames local variable names and function argument names to ordered symbolic notations (e.g., {\mycode var0}, {\mycode var1}) \cite{10.1145/3540250.3549162natgen}.
    \item \textit{\fve (FVE)}~\cite{ijcai2021p0512fie1}: Combines FOE and VOE to eliminate all meaningful identifier names through comprehensive functions and variables renaming.
\end{itemize}

\textbf{Format Features}:
These mutation operators retain all non-blank tokens of the code but eliminate formatting information, such as indentation and line separation.
\begin{itemize}
    \item \textit{\inr (INR)}~\cite{wang-etal-2023-recode}: Removes indentation while preserves the original token sequence.
    \item \textit{\lsr (LSR)}~\cite{jain-etal-2021-contrastive}: Removes line breaks and token separators, concatenating code tokens with space delimiters.
    \item \textit{\tok (TOK)}: Extracts all code tokens and formats them as a JSON list, following the tokenization function provided by CodeSearchNet~\cite{csn}.
\end{itemize}

\textbf{Implementation Details}: These mutation operators eliminate language-specific reserved keywords through natural language conversion and perturb the logic by mutating control flow, e.g., switching {\mycode for} loop into {\mycode while} loop.
\begin{itemize}
    \item \textit{\nl (NL)}: Converts source code to natural language format. Specifically, this mutation operator will parse the source code into an AST, identify programming constructs (e.g., {\mycode for} loops and unary/binary operators) by AST node type, and apply predefined transformation rules that map them to natural language phrases.
    For instance, the ``\(\leq\)'' operator will be expressed as ``{\mycode is less than or equal to}''.
    The integrated transformation rules are disclosed in our artifact.
    \item \textit{\lt (LT)}: Performs equivalent transformations between {\mycode for} and {\mycode while} loop constructs, following the methodology of CodeCleaner~\cite{codecleaner}.
\end{itemize}

Similar to RQ1, we include two baseline setups to reflect the effect of ICL with different mutation operations: \textit{Origin}, in which the original ICL code examples are provided without applying any mutations, and \textit{Zero Shot}.

\newcommand*{\cn}{Concise Name\xspace}
\newcommand*{\vn}{Verbose Name\xspace}
\newcommand*{\vhr}{Variable High-Frequency Replacement\xspace}
\newcommand*{\vmr}{Variable Multi-Word Replacement\xspace}

\subsubsection{Experimental Design for RQ3 on Effect of Code Naming Styles in ICL}

Existing studies have indicated different identifier naming styles as a factor that may affect code comprehension for both human developers~\cite{camel-snake, andrea-descriptive-icpc} and LLMs~\cite{10.1145/3691620.3695072poorcodesumeval, wang-etal-2023-recode}.
To further understand how code naming styles influence ICL performance, we investigate the effectiveness of different identifier naming styles in code demonstrations.
We include the VOE mutation used in RQ2 that yields code incorporating the ordered variable names, and develop four additional operators to synthesize identifier names in different styles.
The design of naming mutation operators considers the fundamental properties of identifier naming in format, information, and length~\cite{10.1145/2652524.2652593, 10.1145/2601248.2601251, 7503707improving_textual, andrea-descriptive-icpc}:
Format refers to the style to which identifier name formats adhere to commonly accepted practices and align with programming language conventions (e.g., snake\_case naming in Python and camelCase naming in Java)~\cite{camel-snake}.
Information refers to the degree to which identifiers are named with words describing their functional intention and meanings.
Length considers the length of identifiers, which represents a critical property of code readability~\cite{8115654}.
Considering these properties, we implement the following mutation operators in addition to VOE for a comprehensive analysis:
\begin{itemize}
    \item \textit{\cn (CN)}: We employ Qwen2.5-Coder-32B-Instruct~\cite{qwen2025qwen25technicalreport} to refactor all local variable names into standardized and concise naming conventions.
    \item \textit{\vn (VN)}:
    We employ Qwen2.5-Coder-32B-Instruct to rename variables using multiple-word descriptors, averaging three words per identifier.
    These identifier names provide detailed functional descriptions for each local variable and function parameter.
    \item \textit{\vhr (VHR)}: Following the existing methods~\cite{10.1145/3691620.3695072poorcodesumeval, 10.1145/3404835.3462840}, we replace original identifiers with high-frequency variable names collected for each programming language. The resulting identifier names conform to common formatting conventions but contain noise on naming information regarding their intended purpose.
    \item \textit{\vmr (VMR)}: We rename original identifiers using combinations of three high-frequency variables, formatted according to language-specific common practice (i.e., snake\_case for Python and camelCase for Java).
\end{itemize}

Qwen2.5-Coder-32B-Instruct is selected for renaming because of its impressive capability on code-to-code translation tasks by effectively identifying the intention of the original source code~\cite{10.1145/3790101}.
It is deployable on our server and shows satisfactory performance on renaming the variables.
In addition, we include the \textit{Zero Shot} and \textit{Origin} configurations from RQ2 as the baseline for comparison.
We conducted an AST structure comparison between the original code and the mutated code to verify that the structural integrity is maintained and that the two versions only differ in identifier names.
Similar to RQ2, the runtime behavior of mutated code is verified to be preserved through executing the tests.

\subsection{Selection of LLMs}
We consider four popular LLMs from three families as our subjects.
For the OpenAI GPT family, we select \gpt~\cite{gpt-4o-mini} as the representative model due to its stable performance and wide availability.
For open source models, we select Qwen2.5-7B-Instruct (\textit{abbr.} \qwensmallabbr) and Qwen2.5-32B-Instruct (\textit{abbr.} \qwenlargeabbr)~\cite{qwen2025qwen25technicalreport} from Qwen and DeepSeek-Coder-V2-Lite-Instruct (\textit{abbr.} \dscoderabbr)~\cite{dscoder} from DeepSeek, as these LLMs are widely studied as representative effective LLMs \cite{ragstudyjy, coderagbench, cui-etal-2024-efficiently, jiang2026surveyLLM} and are deployable on our experimental server infrastructure.

For all LLMs, we adopt \textit{temperature=0} to minimize the randomness of LLMs and \textit{max\_tokens=8000} to prevent the responses from being pruned due to the length limitation.
We design the prompt as shown in \Cref{fig:prompt-template} following the officially provided prompt templates\footnote{LiveCodeBench: \url{https://github.com/LiveCodeBench/LiveCodeBench}; DevEval: \url{https://github.com/seketeam/DevEval}.} for all LLMs, and then apply the same ICL content template to ensure a fair comparison.
We adopt the default maximum context length for all LLMs (128K for \gpt~\cite{gpt-4o-mini} and Qwen models~\cite{qwen2025qwen25technicalreport}; 32K for \dscoderabbr~\cite{dscoder}), which is sufficient to accommodate all the prompts synthesized in our study since the maximum token number is approximately 11K for LeetCode and 27K for DevEval under \(k{=}5\) configuration.
As the context window size of modern LLMs continues to expand~\cite{anthropic-2026, openai-2026-gpt5.5}, the prompt truncation issue will be mitigated and future research can explore the effectiveness of ICL with more examples and longer code snippets.
We expose \(k\) as a configurable parameter in our artifact~\cite{artifact}, facilitating future research to reproduce our results on different \(k\) values.
The responses of \gpt are fetched via its API platform.
The open-source models from Qwen and DeepSeek are powered by the vLLM inference engine~\cite{vLLM} and deployed on a server running \textit{AlmaLinux 9}, equipped with two NVIDIA RTX 6000 Ada Generation GPUs.

\subsection{Performance Metrics}

Following existing works evaluating LLM code generation~\cite{chen2021humaneval, incoder} and common evaluation practices over LiveCodeBench and DevEval,
we use Pass@1 as the evaluation metric for all RQs rather than code similarity scores, as it is more appropriate for problem-solving and code generation tasks.

\section{Experiment Results and Analysis}

In this section, we present the experimental results of the three research questions and discuss the insights behind the observations.

\subsection{RQ1: Effective ICL-Incorporated Information}\label{subsec:RQ1-Result}

\begin{table*}
    \centering
    \smaller
    \setlength{\abovecaptionskip}{2pt}
    \caption{Code Generation Performance (Pass@1) with ICL Examples Retrieved by Different Retrieval Methods (LC-P=LeetCode-Python, LC-J=LeetCode-Java, DE=DevEval).}
    \label{tab:RQ1}
    \begin{tabular}{c|ccc|ccc|ccc|ccc}
    \toprule[1pt]
        \multirow{2}{*}{Retriever} & 
        \multicolumn{3}{c|}{\gpt} & 
        \multicolumn{3}{c|}{\qwensmallabbr} &
        \multicolumn{3}{c|}{\qwenlargeabbr} &
        \multicolumn{3}{c}{\dscoderabbr} \\
        \cmidrule(lr){2-4} \cmidrule(lr){5-7} \cmidrule(lr){8-10} \cmidrule(lr){11-13}
        & LC-P & LC-J & DE & LC-P & LC-J & DE & LC-P & LC-J & DE & LC-P & LC-J & DE\\
    \midrule[1pt]
        GT & 78.45 & 84.81 & 84.51 & 86.46 & 90.06 & 72.95 & 98.62 & 97.24 & 78.77 & 63.54 & 69.06 & 78.70\\
    \midrule
        BM25 & 40.33 & 39.78 & 25.16 & 34.53 & 28.45 & 18.57 & 54.97 & 51.38 & 26.70 & 28.18 & 31.22 & 20.11\\
        gist-large & 40.88 & 38.67 & 37.63 & 35.36 & 32.04 & 24.67 & 52.49 & 52.21 & 36.79 & 26.80 & 27.07 & 23.41\\
        jina-v2 & 39.78 & 39.78 & 36.86 & 37.02 & 30.11 & 23.83 & 56.63 & 52.76 & 35.39 & 27.07 & 28.18 & 24.39\\
        codesage & 41.44 & 39.78 & 37.91 & 33.98 & 28.73 & 24.18 & 53.87 & 53.87 & 35.88 & 27.07 & 27.90 & 26.70\\
    \midrule
        Algo. Tag & 40.88 & 38.67 & - & 31.22 & 29.83 & - & 55.25 & 53.04 & - & 28.18 & 26.24 & -\\
    \midrule
        Random & 38.40 & 40.88 & 16.19 & 32.04 & 30.39 & 12.75 & 53.59 & 50.28 & 16.75 & 27.07 & 27.90 & 13.31\\
    \midrule
        ZS & 39.78 & 41.71 & 14.86 & 38.40 & 33.43 & 11.42 & 55.25 & 54.97 & 15.21 & 30.94 & 34.81 & 12.89\\
    \bottomrule[1pt]
    \end{tabular}
\end{table*}

In this research question, we investigate whether incorporating supportive code examples retrieved under various similarity measures can enhance LLMs' reflection and inference abilities during code generation.
\Cref{tab:RQ1} presents the results of retrieving five relevant documents (k=5).
Our goal is to understand under which retrieval strategies and what ICL example characteristics RAG-ICL benefits or fails to benefit LLM-based code generation.

As shown in \Cref{tab:RQ1}, LLMs achieve superior performance gains under \textit{Ground Truth} configuration on both benchmarks, suggesting that LLMs can recognize and utilize direct answers.
However, they demonstrate different performance under ICL configurations without ground truth.
For LeetCode questions, the studied text-similarity-based (BM25, Embedding Models) and algorithm-based retrieval methods, which represent commonly-used yet relatively simple retrieval baselines, do not yield consistent improvements to LLM code generation.
Specifically, ICL under these configurations does not result in performance improvements compared with \textit{Zero Shot} and \textit{Random} baselines.
To verify this observation, we further conducted statistical significance tests (i.e., Wilcoxon signed-rank test~\cite{wilcoxon-test}, paired \(t\)-test~\cite{Ross2017pairedttest} and Linear Mixed Model~\cite{west2014a}) between Pass@1 under \textit{Zero Shot} and these retrieval configurations with null hypothesis \(H_{0}\): \textit{the median of the differences (Retrieval - Zero Shot) is not greater than zero} and alternative hypothesis \(H_{1}\): \textit{the median of the differences is greater than zero}, as well as Holm-Bonferroni p-value correction~\cite{holmbonferroni} (denoted as \(p_{HB}\)) and Bootstrap Confidence Interval (CI)~\cite{bootstrap-interval}.
The results show that the differences are not statistically significant (\(p{>}0.05\); \(p_{HB}{=}1.0\) and no Bootstrap CI above zero for all models and configurations).
Meanwhile, for DevEval tasks, ICL code examples extracted from the project repository effectively enhance all studied LLMs' Pass@1 performance.
For instance, \qwenlargeabbr achieves 36.79\% with ICL code examples retrieved by gist-large, significantly surpassing 16.75\% under \textit{Random} and 15.21\% under \textit{Zero Shot}.
The statistical significance tests also confirm that the improvements under ICL configurations are statistically significant compared with \textit{Zero Shot} (\(p{<}10^{-10}\); \(p_{HB}{<}10^{-10}\) and all Bootstrap CI above zero for all models and configurations except \textit{Random}).
These observations collectively suggest that the useful information provided by ICL varies across different coding tasks, and the limited improvements observed with the studied retrievers in LeetCode may be attributed to their failure to surface such useful information, as the notable performance gains under GT demonstrate that ICL has the potential to benefit code generation when high-quality examples are available.
Motivated by these observations and aiming to provide insights for retrieval algorithm design and optimization, we further investigate the effective information within ICL examples in \Cref{subsubsec:RQ4.1-deveval} and \Cref{subsubsec:RQ4.1-leetcode}.

\subsubsection{Beneficial Contextual Information in Retrieval Results}\label{subsubsec:RQ4.1-deveval}

\begin{figure}
    \centering
    \includegraphics[width=.9\linewidth]{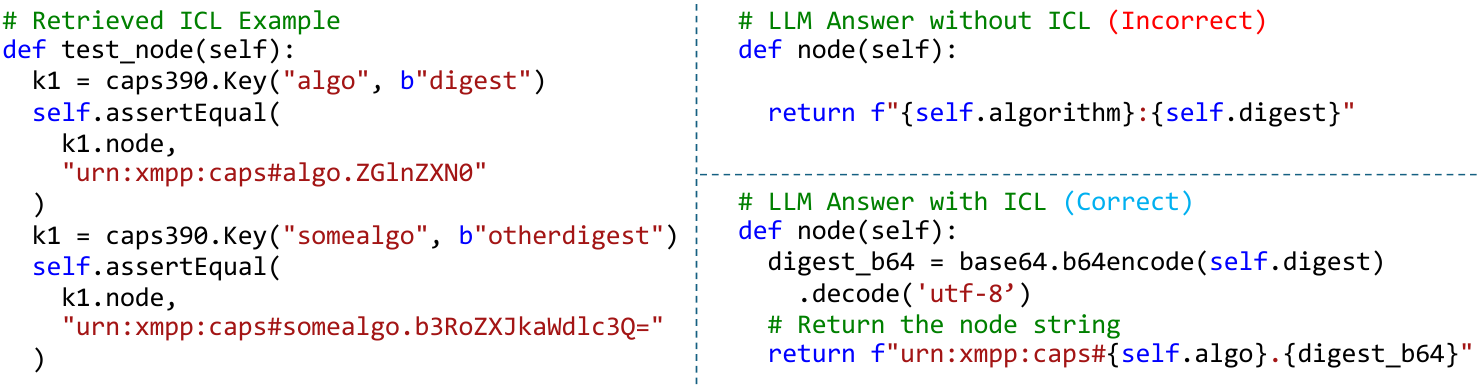}
    \setlength{\abovecaptionskip}{0pt}
    \setlength{\belowcaptionskip}{0pt}
    \caption{A \qwenlargeabbr Failure Case under \textit{Zero Shot} Fixed with ICL-Incorporated Input/Output Demonstration.}
    \label{fig:case-io-demo}
\end{figure}

\begin{figure}
    \centering
    \includegraphics[width=.9\linewidth]{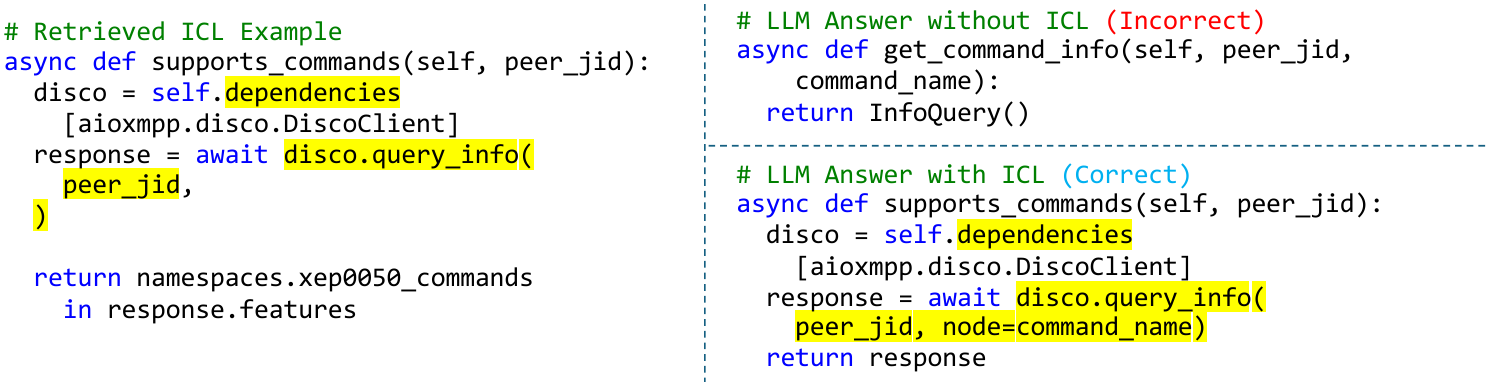}
    \setlength{\abovecaptionskip}{0pt}
    \setlength{\belowcaptionskip}{0pt}
    \caption{A \qwenlargeabbr Failure Case under \textit{Zero Shot} Fixed with ICL-Incorporated Required Function and Class Attribute Information. The related external functions and class attributes are highlighted.}
    \label{fig:case-required-func-var}
\end{figure}

To understand which retrieved content genuinely contributes to code generation, we analyze RAG examples drawn from the current repository using DevEval results.
Through a preliminary case study on randomly sampled failure attempts with ICL, we identify two main categories of RAG-related issues:
\begin{itemize}
    \item \textbf{Missing essential demonstrations:} (1) The LLM infers the general logic correctly but produces incorrect input/output formats (as shown in \Cref{fig:case-io-demo}, where \qwenlargeabbr incorrectly omits the string prefix under \textit{Zero Shot}). (2) The RAG content lacks helper functions or class attributes required by the target function, leaving the LLM unaware of their usage (as shown in \Cref{fig:case-required-func-var}, where \qwenlargeabbr hallucinates a non-existing class under \textit{Zero Shot}, but effectively learns the required functions and class attributes from ICL examples).
    \item \textbf{Namespace-related errors:} (1) The LLM calls valid functions or attributes but omits their necessary namespaces. (2) The RAG content includes functions with identical names as the target, causing the LLM to mistakenly copy them as ground truth.
\end{itemize}
We further analyze both situations, i.e., \textit{I/O demonstration and function/variable recall}, and \textit{repository architecture information}, to identify their roles in performance differences.

\textit{Input/Output Demonstration and Function/Variable Recall.}
We first examine how well RAG content captures input/output behavior and required functions/variables, and how this relates to performance.
Concretely, we compute (1) \textit{unittest recall}, since unittests provide comprehensive, developer-selected I/O demonstrations, and (2) recall of variables and functions/classes required by the ground-truth solution.

We observed that for documents retrieved by BM25, unittest recall is only 2.38\%, whereas the embedding-based retrievers achieve much higher recall, ranging from 51.93\% to 66.92\%.
Similarly, BM25-based RAG covers 48.7\% of variables and 38.9\% of functions/classes used in the ground truth, while embedding models reach 61\%–66\%.
These gaps align with overall performance: embedding-based retrievers consistently outperform BM25 across target LLMs.

To further demonstrate this causal effect, we manually augment the BM25-retrieved RAG with the target unittests and function/variable usages.
Given that such contextual information is lacking in BM25-retrieved examples but present under \textit{Embedding Models} configurations, we hypothesize that augmenting BM25-retrieved examples will result in a performance increase if LLMs can benefit from input/output demonstration and required function/variable names.
Specifically, we manually retrieve the ground truth solution, extract all the variables, functions, and class attributes that appear, then incorporate them into the prompt using JSON format with a brief instruction.
We also randomly retrieve one unittest function marked by DevEval and serve it as an ICL code example.
To avoid LLMs overfitting to visible test cases, we select 586 DevEval tasks with at least two unittests, ensuring the existence of hidden test cases and fair comparison.

\begin{table}
    \centering
    \footnotesize
    \setlength{\abovecaptionskip}{2pt}
    \setlength{\belowcaptionskip}{0pt}
    \caption{Effect of Augmenting BM25 RAG Content with Target Unittests and Required Function \& Variable names. Tasks incorporated with only one unittest are excluded.}
    \label{tab:bm25-enhance}
    \begin{tabular}{lcccc}
    \toprule
        Retriever & \gpt & \qwensmallabbr & \qwenlargeabbr & \dscoderabbr \\
    \midrule
        BM25 & 19.62 & 16.04 & 22.18 & 16.89 \\
        BM25 + unittest/func/var & 27.65 & 20.48 & 28.67 & 23.04 \\
    \midrule
        gist-large & 30.20 & 20.99 & 31.23 & 18.77 \\
        codesage-small & 28.67 & 19.80 & 27.65 & 22.01 \\
    \bottomrule
    \end{tabular}
\end{table}

As shown in \Cref{tab:bm25-enhance}, augmenting BM25 retrievals with unittests and required function/variable information (Row 3) yields substantial performance gains (\(p{<}10^{-10}\); \(p_{HB}{<}10^{-10}\) and all Bootstrap CI above zero for all models), achieving results comparable to embedding-based retrievers.
This finding indicates that the absence of such critical contextual information largely accounts for BM25’s performance gap compared with embedding models.

\textit{Repository Architecture-Related Information.}
Our inspection of failure cases also highlights frequent namespace errors, where LLMs conflate functions with identical names across modules.
To assess the importance of repository structure, we incorporate explicit namespace metadata into RAG content for \textit{BM25}, as well as the top-performing Embedding Model configuration (\textit{gist-large}).
\Cref{tab:namespace} shows that including namespace information consistently improves code generation performance across models.
As shown in \Cref{tab:namespace}, augmenting ICL code examples with namespace information consistently improves code generation performance across all studied models (\(p{<}0.05\); \(p_{HB}{<}0.05\) and all Bootstrap CI above zero), confirming that beyond example content, structural cues such as namespaces and file hierarchies are also essential for effective repository-level code generation.

\begin{table}
    \centering
    \footnotesize
    \setlength{\abovecaptionskip}{2pt}
    \caption{Effect of Incorporating Namespace Information into RAG Content. ``w/o'' and ``w/'' denote without and with namespace information provided as context, respectively.}
    \label{tab:namespace}
    \begin{tabular}{lcccccccc}
    \toprule
        \multirow{2}{*}{Retriever} & \multicolumn{2}{c}{\gpt} & \multicolumn{2}{c}{\qwensmallabbr} & \multicolumn{2}{c}{\qwenlargeabbr} & \multicolumn{2}{c}{\dscoderabbr} \\
    \cmidrule(lr){2-3} \cmidrule(lr){4-5} \cmidrule(lr){6-7} \cmidrule(lr){8-9}
        & \textit{w/o} & \textit{w/} & \textit{w/o} & \textit{w/} & \textit{w/o} & \textit{w/} & \textit{w/o} & \textit{w/} \\
    \midrule
        BM25 & 25.16 & 26.14 & 18.57 & 20.81 & 26.70 & 30.34 & 20.11 & 23.13\\
        gist-large & 37.63 & 39.10 & 24.67 & 27.82 & 36.79 & 39.66 & 23.41 & 26.70\\
    \bottomrule
    \end{tabular}
\end{table}

\subsubsection{Limitations of LLMs in Reflection \& Inference via ICL}\label{subsubsec:RQ4.1-leetcode}
As shown in \Cref{tab:RQ1}, providing similar questions and their corresponding solutions as in-context examples does not facilitate LLMs' code generation on LeetCode problems.
To further understand the ICL effects in this setup, we examine whether LLMs incorporate patterns from the retrieved examples into their generated code.
Specifically, we compute CodeBLEU~\cite{codebleu} similarity scores between LLM outputs and the retrieved ICL examples under two conditions: \textit{with ICL} and \textit{without ICL}.
We then apply the Wilcoxon signed-rank test~\cite{wilcoxon-test}, a standard non-parametric method for comparing paired samples, to assess whether ICL leads to significant increases in similarity between generated code and retrieved examples.
The results reveal no significant difference (\(p \geq 0.99\) across all configurations), indicating that retrieved examples neither meaningfully influence the generation process nor introduce systematic bias, suggesting that although retrieved examples may contain semantically relevant problem-solving logic, LLMs largely fail to extract and apply such patterns during code generation.

In addition, building on the DevEval finding that RAG-provided required functions and variables contribute to LLM performance, we further investigate whether such contextual information is equally beneficial on LeetCode.
Specifically, we augment BM25-retrieved content with the required functions and variables extracted from ground-truth solutions.
Since LeetCode problem descriptions already include input/output demonstrations, we do not additionally incorporate unittests.

\begin{table}[tbp]
    \centering
    \footnotesize
    \setlength{\abovecaptionskip}{2pt}
    \caption{Effect of Incorporating Required Function/Variable Info into RAG Content on LeetCode Questions.}
    \label{tab:func-var-leetcode}
    \begin{tabular}{lcccccccc}
    \toprule
        \multirow{2}{*}{Retriever} & \multicolumn{2}{c}{\gpt} & \multicolumn{2}{c}{\qwensmallabbr} & \multicolumn{2}{c}{\qwenlargeabbr} & \multicolumn{2}{c}{\dscoderabbr} \\
    \cmidrule(lr){2-3} \cmidrule(lr){4-5} \cmidrule(lr){6-7} \cmidrule(lr){8-9}
        & Python & Java & Python & Java & Python & Java & Python & Java\\
    \midrule
        BM25 & 40.33 & 39.78 & 34.53 & 28.45 & 54.97 & 51.38 & 28.18 & 31.22 \\
        BM25 + func/var & 41.44 & 37.85 & 35.36 & 31.22 & 53.59 & 49.17 & 27.35 & 29.56 \\
    \bottomrule
    \end{tabular}
\end{table}

As shown in \Cref{tab:func-var-leetcode}, the results illustrate that the required functions and variables do not enhance LLMs' performance on LeetCode questions (\(p{>}0.1\); \(p_{HB}{>}0.6\); no Bootstrap CI above zero for all models with both Python and Java), although they are helpful in DevEval tasks.
We consider that this is largely due to structural differences between the two benchmarks.
DevEval tasks often require calling functions and using class attributes from the same project, so explicitly providing such information via RAG is beneficial.
In contrast, for the self-contained LeetCode questions, the functions extracted from ground-truth solutions are usually self-defined helpers or standard-library/third-party APIs that LLMs already know, and most extracted variables are local variables whose names can be arbitrarily assigned.
As a result, explicitly providing these functions and variables on LeetCode offers little additional benefit.

In summary, under the simple, widely-adopted retrievers studied in this work, solutions to similar questions do not effectively enhance LLMs' reflection-based reasoning for code synthesis, and required function/variable names, though beneficial for repository-level tasks, offer little advantage in self-contained scenarios.
Comparative results across models further show that weaker LLMs (e.g., \qwensmallabbr) are more easily affected by partially relevant examples.
By contrast, stronger models (e.g., \qwenlargeabbr) exhibit more stable and robust performance across different ICL contexts, indicating their potential ability to filter out irrelevant information.
However, even these relatively powerful models fail to fully leverage truly useful patterns from retrieved code under our studied configurations, revealing a research gap in enhancing both retrieval techniques and LLMs' reflection and inference capabilities.

\begin{tcolorbox}[answerbox]
\textbf{Answer to RQ1}:
The efficiency of ICL varies across different scenarios: incorporating relevant content that clearly specifies input/output behavior, required function/variable information, and repository structure (e.g., namespaces) yields noticeable gains in repository-level tasks, suggesting that fine-grained contextual and architectural signals are crucial for effective RAG-based ICL in code generation. 
By contrast, providing similar question–solution pairs as ICL under the studied retrievers generally fails to enhance LLMs’ reflection-based code generation: across these retrieval methods, performance remains comparable to, or worse than, randomly retrieved examples or no examples scenarios.
This indicates that the simple and straightforward textual-similarity-based retrieval techniques do not reliably surface the problem-solving logic that LLMs need.
\end{tcolorbox}

\subsection{RQ2: Effect of Different Code Features}

\begin{table}
    \centering
    \smaller
    \setlength{\abovecaptionskip}{2pt}
    \caption{Code Generation Performance (Pass@1 in \%) with ICL Examples Mutated by Different Operators on Code (LC-P=LeetCode-Python, LC-J=LeetCode-Java, DE=DevEval).}
    \label{tab:RQ2}
    \begin{tabular}{c|ccc|ccc|ccc|ccc}
    \toprule[1pt]
        \multirow{2}{*}{Mutation} &
        \multicolumn{3}{c|}{\gpt} &
        \multicolumn{3}{c|}{\qwensmallabbr} &
        \multicolumn{3}{c|}{\qwenlargeabbr} &
        \multicolumn{3}{c}{\dscoderabbr}\\
    \cmidrule(lr){2-4}
    \cmidrule(lr){5-7}
    \cmidrule(lr){8-10}
    \cmidrule(lr){11-13}
        & LC-P & LC-J & DE & LC-P & LC-J & DE & LC-P & LC-J & DE & LC-P & LC-J & DE \\
    \midrule[0.75pt]
        Origin & 78.45 & 84.81 & 37.91 & 86.46 & 90.06 & 24.67 & 98.62 & 97.24 & 36.79 & 63.54 & 69.06 & 26.70 \\
    \midrule
        FOE & 69.34 & 80.94 & 31.60 & 82.87 & 88.12 & 22.78 & 96.69 & 97.51 & 31.18 & 55.52 & 65.75 & 19.69 \\
        VOE & 61.88 & 74.59 & 31.74 & 62.98 & 63.54 & 20.25 & 93.09 & 96.41 & 31.04 & 41.44 & 48.34 & 16.33 \\ 
        FVE & 58.29 & 71.27 & 25.58 & 55.52 & 61.33 & 17.66 & 89.23 & 94.48 & 26.91 & 42.54 & 49.72 & 16.26 \\ 
    \midrule
        INR & 77.62 & 87.02 & 35.95 & 83.98 & 89.23 & 22.92 & 94.48 & 97.24 & 35.25 & 64.09 & 73.48 & 23.69 \\ 
        LSR & 72.10 & 83.98 & 34.48 & 81.49 & 85.64 & 21.02 & 93.92 & 97.24 & 33.36 & 73.20 & 80.39 & 23.20 \\ 
        TOK & 67.13 & 72.38 & 29.43 & 71.27 & 79.56 & 20.81 & 87.02 & 93.09 & 32.45 & 79.01 & 85.64 & 22.78 \\
    \midrule
        NL & 79.01 & 79.01 & 34.90 & 72.93 & 61.60 & 19.55 & 92.27 & 93.92 & 32.24 & 66.02 & 66.02 & 22.49 \\ 
        LT & 69.06 & 76.52 & 36.65 & 85.08 & 79.28 & 21.02 & 97.51 & 95.30 & 34.83 & 59.94 & 60.22 & 24.25 \\
    \midrule
        ZS & 39.78 & 41.71 & 14.86 & 38.40 & 33.43 & 11.42 & 55.25 & 54.97 & 15.21 & 30.94 & 34.81 & 12.89 \\
    \bottomrule[1pt]
    \end{tabular}
\end{table}

This research question investigates the impact of various code features on code snippets within the context of ICL for code generation tasks.
\Cref{tab:RQ2} presents the results with different mutation operators applied to ICL code examples.
For LeetCode, we apply mutation operators on the \textit{Ground Truth} since the similar questions retrieved by popular retrievers cannot help enhance the code generation performance as observed in RQ1.
This configuration provides a clean setup to mitigate the effect of retrieved document quality, thus serving as a ground to investigate the upper bound of ICL effectiveness on contest-level tasks~\cite{coderagbench}.
For DevEval, we use code snippets from the current repository retrieved by embedding models as the original ICL content.
We select the embedding model achieving the highest performance for each LLM (\textit{gist-large} for \qwensmallabbr/\qwenlargeabbr, and \textit{codesage-small} for \gpt/\dscoderabbr).
This setup represents a common ICL-assisted code generation scenario by using code snippets present prior to the target functions.
These ICL setups collectively provide the opportunity to investigate the effectiveness of code features in different directions.
We analyze the results from three perspectives.

\textbf{Impact of Different Mutation Operators.}
Our analysis reveals distinct patterns in how different code features affect LLM performance, verified by the same statistical significance testing procedure described in \Cref{subsec:RQ1-Result} comparing Pass@1 under \textit{Origin} and each mutation configuration with \(H_0\): \textit{the median of the differences (Mutation - Origin) is not lower than zero}.
Across all mutation operators spanning the three categories, identifier-related code obfuscations (VOE and FVE) consistently induce the most severe performance degradation for both benchmarks.

\begin{figure}
    \centering
    \includegraphics[width=.9\linewidth]{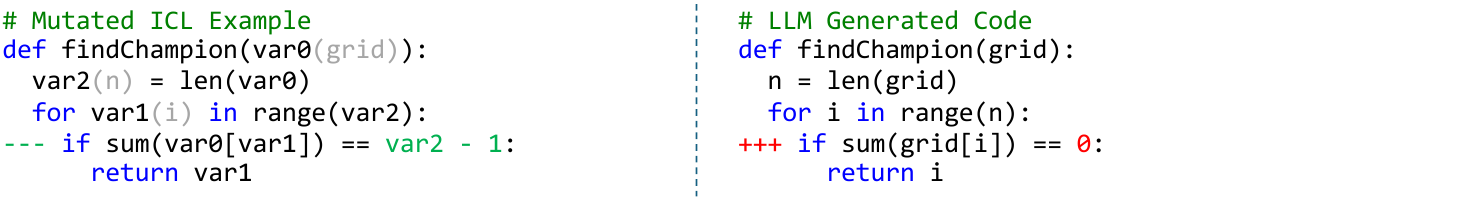}
    \setlength{\abovecaptionskip}{2pt}
    \setlength{\belowcaptionskip}{2pt}
    \caption{An Incorrect Python Code Generated by \gpt under VOE Mutation. Variable names before mutation are in gray; code causing semantic differences is highlighted using code changes.}
    \label{fig:case-voe}
\end{figure}

On the self-contained LeetCode questions, the degradation caused by VOE and FVE is particularly pronounced, with \qwensmallabbr subject to a performance decrease of 30 percentage points under FVE.
Even for stronger models like \gpt, variable name elimination can also cause incorrect semantic interpretation, resulting in a slight but failure-inducing difference, as demonstrated in \Cref{fig:case-voe} and the significance tests (\(p{<}10^{-3}\); \(p_{HB}{<}10^{-3}\); all Bootstrap CI below zero for all models under both VOE and FVE except \qwenlargeabbr with Java).
For DevEval tasks depending on external context, although the absolute performance is lower due to task difficulty, identifier obfuscation remains the dominant failure factor.
For instance, under FVE, DevEval performance drops from 37.91\% to 25.58\% for \gpt, 24.67\% to 17.66\% for \qwensmallabbr, and 36.79\% to 26.91\% for \qwenlargeabbr.
Additionally, the statistical tests confirm the significance of the performance drops under FOE, VOE and FVE compared with \textit{Origin} (\(p{<}0.04\); \(p_{HB}{<}0.04\) and all Bootstrap CI below zero for all models).
These consistent declines suggest that identifier naming information is a key feature for ICL under different relevance of ICL content, implying that ICL prompt designers should adopt code examples with clear, human-readable identifiers describing their intention and functionality.

Among all mutation operators on identifier names, FOE configuration causes minimal performance degradation on LeetCode, compared with a relatively significant decrease on DevEval.
The reason may be that LeetCode solutions mainly depend on common library APIs and user-defined functions whose bodies are present in the same code example.
In contrast, code snippets extracted from DevEval repositories contain more helper functions defined externally and not visible in ICL examples, making LLMs hard to reason the intention of these functions.

\begin{figure}
    \centering
    \includegraphics[width=.9\linewidth]{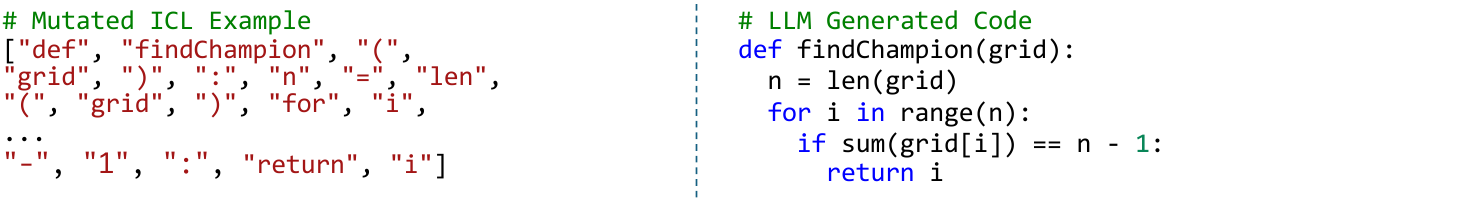}
    \setlength{\abovecaptionskip}{2pt}
    \setlength{\belowcaptionskip}{2pt}
    \caption{A Correct Python Code (Right) Generated by \gpt under Code Mutated with TOK (Left).}
    \label{fig:case-tok}
\end{figure}

\textbf{Impacts on Different Programming Languages.}
Cross-language comparison on LeetCode reveals different behaviors of LLMs in the generation of Python and Java programs, respectively.
In general, under the \textit{Zero Shot} baseline, Java and Python demonstrate comparable performance.
However, under \textit{Ground Truth} configuration, LLMs generally generate more correct Java programs compared with Python (except the more powerful \qwenlargeabbr shows a comparably good code generation performance in both languages), indicating a clear programming language preference on code comprehension and generation.

Moreover, we observe that LLMs exhibit greater sensitivity to format elimination in generating Java programs with ICL.
This is intriguing since Java code is independent of several format changes (e.g., indentation, line separation), while LLMs fail to learn from Java example code with irregular formats.
Meanwhile, Python code, which is sensitive to the format changes while more familiar to LLMs \cite{wang-etal-2023-recode}, can be effectively leveraged by LLMs despite the semantics of the ICL code example being altered due to the format changes.
For example, the TOK mutation results in up to a 12-point drop in Java performance with \gpt.
This exceeds the maximum drop caused by format elimination observed in Python, where LLMs can generally recover the correct code format from extracted tokens, as shown in \Cref{fig:case-tok}.
Similarly, implementation mutations such as NL reduce performance by up to 28 percentage points for (\qwensmallabbr, LeetCode-Java).
The results suggest that LLMs may not access the semantics of the ICL code example by carefully parsing the code based on its format. 
LLMs can defend against the semantic noise caused by incorrect format for the familiar Python language, while struggling with the correct code with irregular format in Java, which is relatively less familiar to LLMs as reported by previous works~\cite{twist2025llmslovepythonstudy, tong-zhang-2024-codejudge}.

\textbf{Impact on Different LLMs.}
Different LLMs exhibit distinct behavioral patterns under various mutation conditions. 
\qwenlargeabbr demonstrates remarkable robustness across all mutation operators, maintaining performance above 87\% for all configurations in LeetCode, and above 30\% in nearly all DevEval configurations except FVE.
This stability suggests superior generalization capabilities and reduced dependence on specific code features of the powerful LLM for both contest-level and repository-level code generation tasks.
Conversely, weaker LLMs such as \qwensmallabbr show more pronounced performance degradation as information is progressively removed from the ICL context.
The pattern indicates a correlation between the LLMs' coding ability (demonstrated by the performance on code generation benchmarks~\cite{chen2021humaneval, zhuo2025bigcodebench}) and context code leveraging with information elimination.
In addition, we observe that \dscoderabbr presents a unique behavioral profile on LeetCode: its performance significantly lags behind other models under \textit{Origin} configuration, while it becomes competitive under format and implementation detail mutation configurations.
This observation may suggest that different LLMs may prefer unique representation formats when processing source code in context.

Our findings bring implications for the design of ICL-based prompting strategies and evaluation benchmarks.
Prompt designers are advised to adopt code examples with clear, human-readable identifiers that convey intent and functionality.
Moreover, benchmark datasets that aim to assess the real-world applicability of LLMs should carefully preserve semantic fidelity in code samples.
For code with anonymous naming styles, developers are advised to refactor the code to facilitate LLMs' code comprehension.

\begin{tcolorbox}[answerbox]
\textbf{Answer to RQ2}:
Meaningful identifier names are more influential on effective ICL in code generation for both contest-level and repository-level tasks, compared with formats and implementation details.
LLMs show greater sensitivity to format and implementation mutations in Java than Python, while larger or specialized models (e.g., \qwenlargeabbr, \dscoderabbr) are more robust to feature elimination.
These results collectively yield insights on ICL code example construction regarding the code features which developers should prioritize, e.g., clear identifier semantics.
\end{tcolorbox}

\subsection{RQ3: Effect of Different Code Naming Styles}
Prior works indicated that identifier names may largely affect code comprehension for both humans and LLMs~\cite{7503707improving_textual, 5714430}.
Our RQ2 results also suggest that perturbing identifiers (VOE, FVE) can substantially affect ICL performance.
Motivated by these findings, RQ3 investigates how different identifier naming styles in ICL demonstrations influence code generation.
\Cref{tab:RQ3} reports Pass@1 under all naming-style configurations.
We analyze the results from three complementary aspects: semantic informativeness, naming convention, and identifier length.

\begin{table*}
    \centering
    \smaller
    \setlength{\abovecaptionskip}{0pt}
    \setlength{\belowcaptionskip}{0pt}
    \caption{Code Generation Performance (Pass@1) with ICL Examples Using Different Naming Styles in Code (LC-P=LeetCode-Python, LC-J=LeetCode-Java, DE=DevEval)}
    \label{tab:RQ3}
    \begin{tabular}{c|ccc|ccc|ccc|ccc}
    \toprule[1pt]
        Mutation & 
        \multicolumn{3}{c|}{\gpt} & 
        \multicolumn{3}{c|}{\qwensmallabbr} &
        \multicolumn{3}{c|}{\qwenlargeabbr} &
        \multicolumn{3}{c}{\dscoderabbr} \\
        \cmidrule(lr){2-4} \cmidrule(lr){5-7} \cmidrule(lr){8-10} \cmidrule(lr){11-13}
        & LC-P & LC-J & DE & LC-P & LC-J & DE & LC-P & LC-J & DE & LC-P & LC-J & DE \\
    \midrule[1pt]
        Origin & 78.45 & 84.81 & 37.91 & 86.46 & 90.06 & 24.67 & 98.62 & 97.24 & 36.79 & 63.54 & 69.06 & 26.70 \\
    \midrule
        CN & 77.90 & 85.36 & 27.89 & 86.19 & 89.78 & 17.52 & 98.62 & 97.79 & 27.05 & 68.23 & 73.48 & 11.07 \\
        VN & 70.44 & 77.62 & 33.57 & 81.22 & 87.29 & 19.55 & 96.41 & 98.34 & 29.78 & 54.42 & 68.51 & 17.03 \\
    \midrule
        VOE & 61.88 & 74.59 & 31.74 & 62.98 & 63.54 & 20.25 & 93.09 & 96.41 & 31.04 & 41.44 & 48.34 & 16.33 \\
    \midrule
        VHR & 56.08 & 58.29 & 23.27 & 60.50 & 65.47 & 16.05 & 85.64 & 88.40 & 22.78 & 37.29 & 37.85 & 10.37 \\
        VMR & 53.59 & 62.43 & 26.84 & 56.08 & 56.63 & 18.08 & 84.25 & 90.88 & 27.33 & 35.36 & 39.50 & 14.86 \\
    \midrule
        ZS & 39.78 & 41.71 & 14.86 & 38.40 & 33.43 & 11.42 & 55.25 & 54.97 & 15.21 & 30.94 & 34.81 & 12.89 \\
    \bottomrule[1pt]
    \end{tabular}
\end{table*}

\begin{figure}
    \centering
    \includegraphics[width=.9\linewidth]{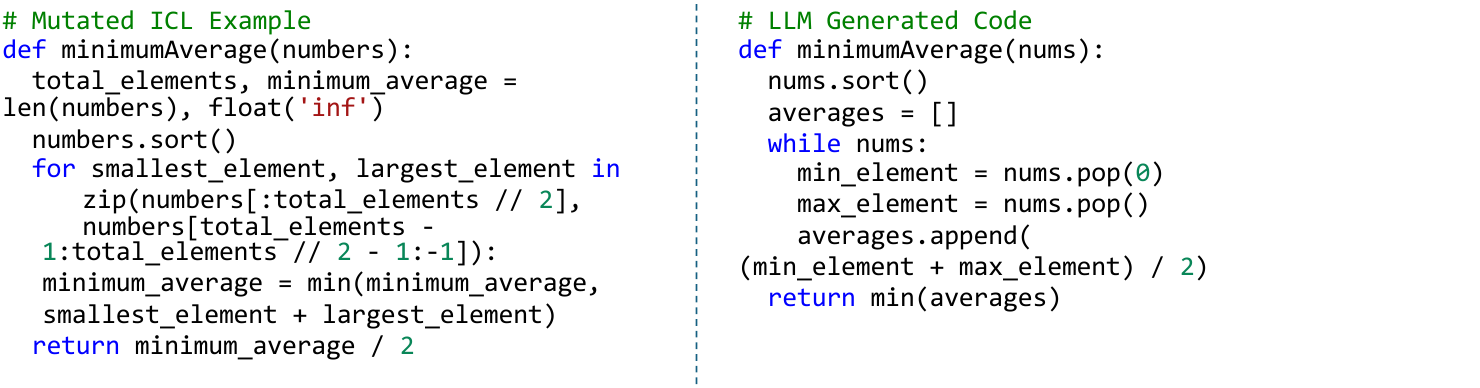}
    \setlength{\abovecaptionskip}{0pt}
    \setlength{\belowcaptionskip}{-2pt}
    \caption{An Incorrect Python Code Generated by \qwensmallabbr under VN Mutation.}
    \label{fig:case-vn}
\end{figure}

\textbf{Impact of Semantic Information in Identifier Names:}
We first compare the ICL code generation performance under the mutations that preserve semantic intent by renaming variables with meaningful descriptors (CN/VN) and the mutations that eliminate useful semantic intent in identifier names (VHR/VMR).
For both LeetCode questions and DevEval tasks, CN and VN yield substantially stronger performance than VHR and VMR across four studied LLMs.
In contrast, VHR and VMR, which replace identifiers with high-frequency names and their combinations, exhibit the largest degradation (\(p{<}10^{-3}\); \(p_{HB}{<}10^{-3}\) and all Bootstrap CI below zero under all combinations of models and mutations) and are generally even worse than VOE.
The results indicate that LLMs still heavily depend on the semantic descriptions in identifiers when learning from ICL demonstrations, regardless of whether the coding task is dependent on external context (DevEval) or not (LeetCode).

\textbf{Impact of Different Naming Conventions:} 
We also examine LLMs' preference for ICL examples using different naming conventions.
Specifically, the variable names in the VOE configuration follow a non-standard naming pattern (e.g., {\mycode var0}, {\mycode var1}), while the variable names in VHR and VMR configurations follow popular snake\_case and camelCase conventions.
However, we observe performance degradation in VHR and VMR configurations compared with VOE on both LeetCode questions and DevEval tasks.
The finding suggests that the unfamiliar formatting conventions related to text style are unlikely to hinder LLMs from recognizing helpful code. 
Instead, as discussed previously, the description of variables is more crucial to unlock LLMs' capabilities in recognizing useful ICL code examples.

\textbf{Impact of Identifier Length:}
Finally, we compare CN (relatively concise names) and VN (more verbose, composite names with around three words per identifier).
On LeetCode, CN generally outperforms VN for Python across all models, while the preference is weaker for Java (e.g., \qwenlarge slightly favors VN).
On DevEval, we observe the opposite trend: CN is consistently lower than VN across all four LLMs.
To further understand this discrepancy, we measure the textual similarity between variable names in the \textit{Origin} Python code and those produced by LLM based on the average cosine similarity between BERT embeddings of identifier strings~\cite{devlin-etal-2019-bert}.
For LeetCode Python, CN achieves a higher average similarity than VN (CN: 0.70 vs. VN: 0.45).
For DevEval code examples retrieved by RAG, VN is slightly closer to the \textit{Origin} identifiers than CN (CN: 0.46 vs. VN: 0.52 for both \textit{gist-large} and \textit{codesage-small}).
This discrepancy likely stems from differing naming conventions across task types: contest-style code typically favors concise, standardized single-word identifiers (e.g., {\mycode dp}, {\mycode count}), whereas repository code tends to use multi-word names that fully describe variable intent and distinguish them from other symbols in the codebase.
These similarity scores align with the observed performance differences, suggesting that length alone is not the key contributor to ICL performance.
As demonstrated in \Cref{fig:case-vn}, verbose variable names may distract the model and prevent it from recognizing the usefulness of the code example.
Instead, identifier names that better match the intended semantics (and thus are closer to the \textit{Origin} identifiers) more reliably support ICL.

Experiments in this RQ reveal that identifier naming style plays a critical role in the effectiveness of ICL.
Therefore, we recommend that developers prioritize the use of descriptive, consistent, and semantically meaningful identifiers in their source code when working within AI-integrated development environments (IDEs), in order to improve retrieval quality and enhance the downstream performance of LLM-based assistants.

\begin{tcolorbox}[answerbox]
\textbf{Answer to RQ3}: 
LLMs benefit from ICL demonstrations primarily when identifier names convey meaningful and reliable semantic descriptions.
Meanwhile, standardizing naming conventions or making identifiers longer/shorter has secondary effects compared to semantic informativeness.
\end{tcolorbox}

\section{Discussion}

\subsection{Statistical Assumptions Validation}
We checked statistical assumptions for each paired comparison cell across all RQs.
A cell denotes one fixed comparison setting, e.g., one model and one benchmark slice under one retrieval or mutation configuration.
Table~\ref{tab:rq1-assumptions} reports rejected cells at $\alpha{=}0.05$.
Since the outcomes are binary, the parametric normality assumptions of the paired $t$-test and the Linear Mixed Model (LMM) are violated almost universally, and the symmetry and homoscedasticity assumptions are also frequently rejected.
However, a rejection of the symmetry assumption does not by itself invalidate the Wilcoxon test, as it only needs to hold under $H_0$.
We therefore ground the conclusions on the Wilcoxon signed-rank test together with the Bootstrap CI on the mean paired difference and report the paired $t$-test and LMM only as concordance checks.

\begin{table*}[tbp]
  \centering
  \footnotesize
  \setlength{\abovecaptionskip}{0pt}
  \setlength{\belowcaptionskip}{0pt}
  \caption{Cells Rejecting Each Assumption at $\alpha{=}0.05$ (rejected/total; LC=LeetCode, DE=DevEval).}
  \label{tab:rq1-assumptions}
  \begin{tabular}{llcccccc}
    \toprule
    \multirow{2}{*}{Assumption (test)} & \multirow{2}{*}{Required by} & \multicolumn{2}{c}{RQ1} & \multicolumn{2}{c}{RQ2} & \multicolumn{2}{c}{RQ3} \\
    \cmidrule(lr){3-4} \cmidrule(lr){5-6} \cmidrule(lr){7-8}
    & & LC & DE & LC & DE & LC & DE \\
    \midrule
    Normality of diffs.\ (Shapiro--Wilk~\cite{10.1093/biomet/52.3-4.591}) & Paired $t$-test & 40/40 & 20/20 & 64/64 & 32/32 & 32/32 & 16/16 \\
    Symmetry of diffs.\ (D'Agostino skew~\cite{10.1093/biomet/57.3.679}) & Wilcoxon & 24/40 & 20/20 & 50/64 & 32/32 & 20/32 & 16/16 \\
    Normality of LMM residuals & LMM & 40/40 & 20/20 & 64/64 & 32/32 & 32/32 & 16/16 \\
    Homoscedasticity (Levene~\cite{levene1960robust}) & LMM & 0/40 & 16/20 & 26/64 & 7/32 & 14/32 & 13/16 \\
    \bottomrule
  \end{tabular}

\end{table*}

\subsection{Threats to Validity}

\textit{Representativeness of Mutation Operators.}
This study investigates the contributions of each code feature to effective ICL by eliminating such features by applying mutation operators to the original solution code. 
While we have covered major categories of code features, there may be other aspects of code that could influence ICL effectiveness.
To mitigate this threat, we refer to prior works focusing on code readability, which propose clear definitions regarding factors contained in code snippets~\cite{10.1145/1985441.1985454, learning_metric_code_readability, 7503707improving_textual}, systematically categorize code features into three fundamental dimensions (semantics, format, and implementation details) based on established literature~\cite{10.1145/3540250.3549162natgen, 10.1145/3691620.3695072poorcodesumeval, codecleaner, 10.1145/3428293foe, ijcai2021p0512fie1}, and design mutation operators that comprehensively cover each dimension.

\textit{Representativeness of Studied Code Generation Tasks.}
We adopt two code generation benchmarks in this study.
While these benchmarks cover diverse programming scenarios: algorithmic reasoning (LeetCode) and repository-level development (DevEval), they may not fully represent all real-world software development scenarios.
Additionally, potential data contamination issues may arise from the exposure of benchmarks in the training process of LLMs.
For the representativeness concern, we argue that LeetCode and DevEval are among the most widely adopted benchmarks for evaluating LLM code generation capabilities~\cite{seed2025seedcoderletcodemodel, hui2024qwen25codertechnicalreport, coderag-jiali}, covering diverse algorithmic patterns and software development specifications.
Furthermore, we extend Python code generation tasks in LiveCodeBench to Java to strengthen the generalizability of our findings.
For the data contamination concern, we confirm that the knowledge cutoff dates of all four studied models precede 2024.
Therefore, only a limited subset of LiveCodeBench-v5 (May 2023 to January 2025) and no DevEval tasks (May 2024) may have been included in the training data.

To further substantiate that our findings are not driven by potential data contamination, we partitioned the LeetCode questions into two subsets for each model according to the knowledge cutoff date of each LLM: questions released \textit{before} the cutoff (potentially seen during pre-training) and \textit{after} the cutoff (guaranteed unseen).
We then computed pass@1 of every evaluation scenario in our study on the two subsets separately, and measured the Spearman rank correlation~\cite{zar2005spearman} of the per-scenario pass@1 among \textit{before}-cutoff, \textit{after}-cutoff, and the \textit{overall} full question set.
The correlations are consistently high across all four studied LLMs ($p{<}10^{-18}$ for all paired comparisons), indicating that the relative effectiveness of different ICL configurations is preserved regardless of whether the questions could have appeared in the training data.
This confirms that our conclusions about the contributions of code features to ICL effectiveness are robust to potential data contamination.

\section{Related Work}

We summarize previous works related to this study as follows:

\subsection{Advancements in In-Context Learning}
ICL emerges as a promising approach to guide LLMs to follow instructions and learn from examples.
Research works optimize the effectiveness of ICL through designing Chain-of-Thought prompting strategy \cite{cot-prompting}, building appropriate Reinforcement Learning algorithms \cite{zhang-etal-2022-active}, retrieving helpful examples as context \cite{acecoder}, and using LLMs to indicate effective examples \cite{10.1145/3715908LAIL}.
Besides, some studies focus on investigating the factors affecting ICL performance.
\citet{liu-etal-2022-makes} report that GPT-3's ICL performance highly relies on the selected examples, while examples semantically similar to the original query enhance the overall performance.
\citet{min-etal-2022-rethinking} explore the effect of labeling and point out that labels only have a marginal effect on the overall performance.
\citet{10.1109/ASE56229.2023.00109goodICLdemo} conduct a comprehensive exploration of various factors affecting ICL performance on code intelligence tasks, such as example selection, context arrangement, and example quantities.

Although existing works advance in optimizing ICL for various tasks including coding, they consider code examples as monotonic units and do not explore the contributing features in a code snippet.
This study bridges this gap by systematically investigating the effect of different code features in retrieved code examples, explaining the effectiveness of ICL on coding tasks.

\vspace{-3pt}

\subsection{Study on LLMs' Preference on Code Snippets}
Beyond ICL scenarios, existing works reveal that LLMs show certain preferences for code snippets with similar functionality but unique semantics or formats.
\citet{wang-etal-2023-recode} find that LLMs show different robustness to perturbations on original code snippets in code completion tasks.
\citet{10.1145/3691620.3695072poorcodesumeval} find that current models are vulnerable to poor-readability code.
\citet{codepromptzip} rank the removal priorities of different code token types and train a compressor to achieve flexible compression of code examples embedded into RAG workflow.
In addition, some works enhance LLM performance by utilizing different preferences or enhancing LLM robustness against different codes with similar functionality.
\citet{jain-etal-2021-contrastive} reveal that LLMs are sensitive to semantic-preserving edits and design a contrastive pre-training task, which trains the model to learn the code functionality rather than form.
\citet{wang-etal-2021-codet5} propose CodeT5 based on the architecture of T5~\cite{10.5555/3455716.3455856t5}, leveraging developer-assigned identifier names for effective code comprehension and code completion.
\textsc{NatGen}~\cite{10.1145/3540250.3549162natgen} proposes a novel pre-training objective that aims to recover syntactically uncommon code to its semantically equivalent, naturalized counterpart.
\citet{9825895} provide a novel training method targeting optimization of model robustness against semantics-preserving code transformations.

These optimization techniques are mainly driven by machine learning approaches, lacking explanations on preference of LLM or predictable results before the actual generation.
In comparison, our study systematically mutates different features to understand which properties inside code snippets make the most significant contribution to ICL.
It reveals that LLMs rely more on textual semantic features than on format information and problem-solving logic, providing insights for future pre-training objectives and optimization techniques for LLM-based code intelligence tasks.

\section{Conclusion}

In this paper, we investigate a crucial yet unexplored question: \textit{Which specific properties in ICL code examples contribute significantly to the observed performance improvements?}
Through systematic empirical studies and controlled ablations,
we identify the essential contextual information (required functions, variables, meaningful identifier names) in ICL-assisted code generation.
Despite the overall success of ICL across diverse scenarios, current LLMs remain limited in their ability to generalize problem-solving insights based on reflection and inference.
Our findings provide actionable guidance for designing ICL systems: practitioners should prioritize including relevant context and descriptive identifiers in code examples, while future research should focus on enhancing LLMs' capacity for deeper reasoning and abstraction in programming logic.

\section*{Data Availability}
We make our artifact, including the experiment results and the replication code, publicly available at \url{https://doi.org/10.6084/m9.figshare.32835674} to facilitate follow-up research studies.

\begin{acks}
We would like to thank the anonymous reviewers for their insightful comments.
This work is partially supported by the National Natural Science Foundation of China (Grant No. 92582201), the Hong Kong SAR Research Grant Council/General Research Fund (Ref No. 16206524), and the Hong Kong SAR Research Grant Council/Theme-based Research Scheme (Ref No. T41-517/25-N).
\end{acks}

\bibliographystyle{ACM-Reference-Format}
\bibliography{reference}

\end{document}